\documentclass[twocolumn,preprintnumbers,amsmath,amssymb]{revtex4}
\usepackage{txfonts}
\usepackage{amsmath}
\usepackage{stmaryrd}
\usepackage{amssymb}
\usepackage{multirow}
\usepackage{cases}
\usepackage[none]{hyphenat}

\usepackage{graphicx}
\usepackage{dcolumn}
\usepackage{bm}
\usepackage{braket} 
\usepackage{color}

\def\beq{\begin{equation}}
\def\eeq{\end{equation}}
\def\beqa{\begin{eqnarray}}
\def\eeqa{\end{eqnarray}}
\newcommand{\eq}[1]{Eq.(#1)}
\newcommand{\eqs}[1]{Eqs.(#1)}
\newcommand{\fig}[1]{Fig.(#1)}

\begin{document}
\title{Quantum memory with a single two-level atom in a half cavity}
\author{Yimin~Wang$^1$, Ji\v{r}\'{i}~Min\'{a}\v{r}$^1$, Gabriel~H\'{e}tet$^2$, and Valerio~Scarani$^{1,3}$}
\affiliation{$^1$Centre for Quantum Technologies, National University of Singapore, Singapore\\
$^2$Institute for Experimental Physics, University of Innsbruck, A-6020 Innsbruck, Austria\\
$^3$Department of Physics, National University of Singapore, Singapore}


\begin{abstract}

We propose a setup for quantum memory based on a single two-level atom in a half cavity with a moving mirror. We show that various temporal shapes of incident photon can be efficiently stored and readout by shaping the time-dependent decay rate $\gamma(t)$ between the atom and the light. This is achieved uniquely by an appropriate motion of the mirror without the need for additional control laser or atomic level. We present an analytical expression for the efficiency of the process and study its dependence on the ratio between the incident light field bandwidth and the atomic decay rate. We discuss possible implementations and experimental issues, particularly for a single atom or ion in a half cavity quantum optical setup as well as a superconducting qubit in the context of circuit QED.

\end{abstract}


\maketitle


\section{Introduction}

Efficient and faithful storage of quantum states of light lies at the heart of long distance quantum communication \cite{Duan_2001,Sangouard_2009b} and remarkable progress and achievements have been done in recent years \cite{Hammerer_2010,Simon_2010}. A device allowing such storage, the so-called quantum memory, can be implemented in various physical systems, which can be either atomic ensembles \cite{Hammerer_2010} or single atom like systems (typically single atoms or ions \cite{Specht_2011, Stute_2011}, quantum dots \cite{Yilmaz_2010}, superconducting qubits \cite{Devoret_2004} or NV centers \cite{Dutt_2007}). In the present article we concentrate on the latter situation and consider single atom systems, where much experimental progress has been done recently. The demanding part in these systems is that a strong coupling between the atom and the light is required. In the atomic system, this can be achieved by using high numerical aperture optical elements \cite{Tey_2008, Tey_2009, Chen_2011} or a high finesse cavity \cite{Specht_2011}, where the quantum memory application has already been demonstrated using a mapping of the polarization of the light qubit onto a single  $^{87}{\rm Rb}$ atom. Single atom systems are also well suited for creating and manipulating the quantum information experiments demonstrating entanglement generation between two individual atoms \cite{Wilk_2010} and quantum gate operations between neutral atoms \cite{Isenhower_2010} and ions \cite{Benhelm_2008} have been realized. Moreover, for long lived information storage, one usually needs to transfer the optical coherence into the coherence between ground states. This is usually achieved using another strong laser beam between the excited state and the state used for storage.

In this paper, we propose a quantum memory setup consisting of a single two-level atom in a half cavity, in which we allow for an arbitrary motion of the mirror to modify the atom-light interaction --- a natural extension of the previous work done by other authors \cite{Eschner_2001, Dorner_2002, Wilson_2003, Glaetzle_2010} and \cite{Green_2011}. We show by explicit calculation that various temporal shapes of the input single photon pulse can be efficiently stored by the atom-mirror system, provided the motion of the mirror is optimized. A feature of this scheme is that there is no need for an additional atomic level nor the strong control/transfer laser. We discuss the memory efficiency and fidelity as well as possible implementations, such as a single atom/ion in a half cavity or a superconducting qubit coupled to a 1D transmission line terminated by a SQUID.

The paper is organized as follows: we present a derivation of the optimized time-dependent decay rate that maximizes the efficiency of the storage in Sec. \ref{sec QM model}. We illustrate these results with an example of an input single photon time-bin qubit and discuss possible experimental realizations of the quantum memory scheme  in Sec. \ref{sec Sim_Imp}.\\
~\\

\section{The quantum memory model}
\label{sec QM model}


\subsection{General optical Bloch equations}
\label{sec obe}
\begin{figure}[b]
\includegraphics[scale=0.35]{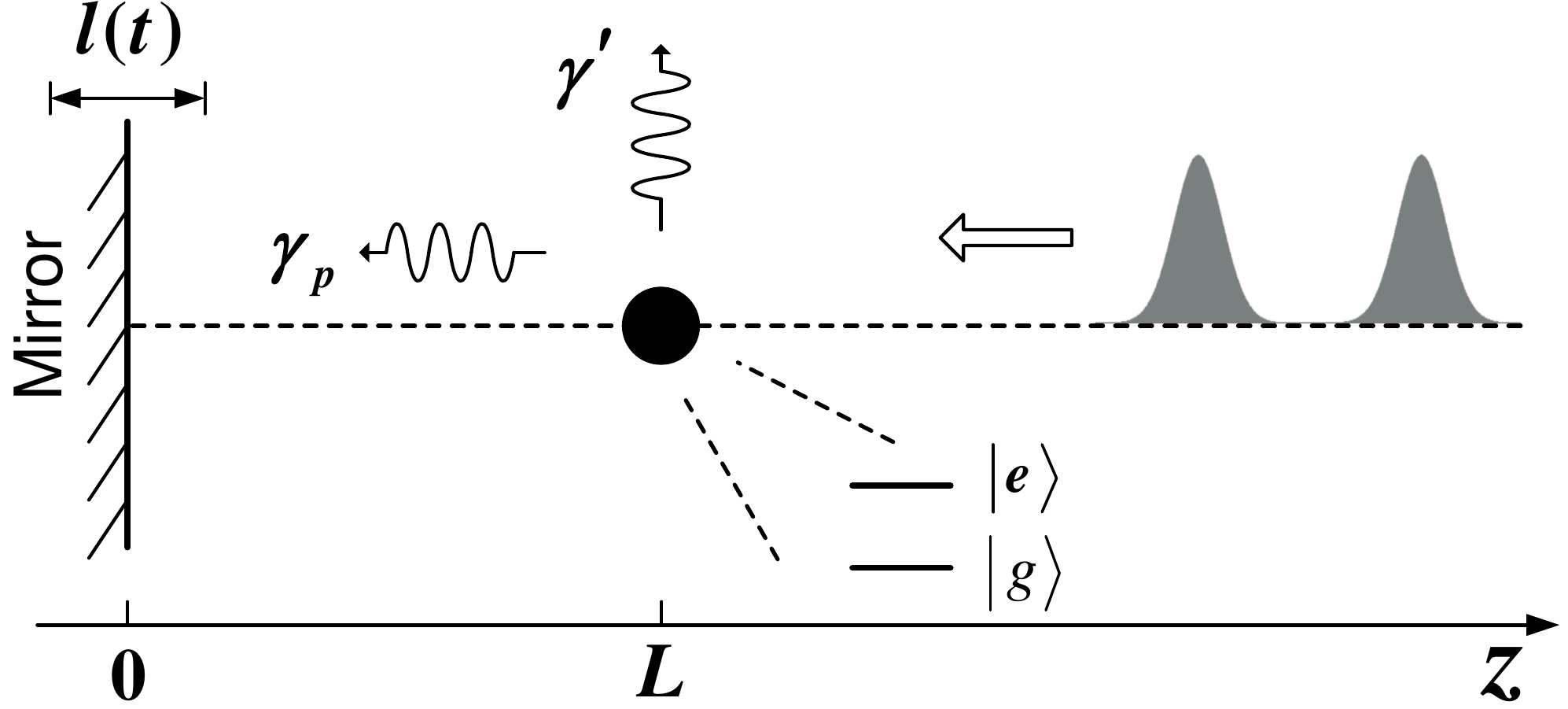}
\caption{ Sketch of the quantum memory setup: an arbitrary single photon wave packet interacts with a two-level atom which has a initial distance $L$ from the movable mirror, whose motion is described by $l(t)$. $\gamma_p$ and $\gamma'$ describe the decay rates into the pulse mode and the environment, respectively (remark: in the implementation that we consider, the pulse durations are much longer than $L/c$).}
\label{fig setup}
\end{figure}

We study a single two-level atom sitting in front of a moving mirror (see \fig{\ref{fig setup}}). The incident pulse propagates along the $z-$axis and first interacts with the atom. The positive frequency part of the continuum electric field operator in the standing wave basis and the interaction picture reads \cite{Blow_1990, Dorner_2002, Domokos_2002}
\beq
    \label{eq E+}
	\hat{\bm E}^{(+)}(z,t)=i \sum_{\lambda} \int_0^{\infty} {\rm d} \omega \,\, A(\omega) {\bm \epsilon}_{\omega,\lambda} \sin (kz) \,e^{-i\omega t} \hat{a}_{\omega}(t),
\eeq
\noindent where $k=\omega/c$, $c$ is the vacuum speed of light, ${\bm \epsilon}_{\omega,\lambda}$ with $\lambda=\{1, 2\}$ denotes the unit polarization of mode $\omega$ and the coefficient $A(\omega)$ accounts for the correct normalization of the electric field (i.e. that the total energy of a single photon Fock state of frequency $\omega_0$ is $\hbar \omega_0$). We denote the initial distance between the atom and the mirror by $L$. The main goal of this article is to investigate the dynamics of a two-level atom and a pulse in front of a moving mirror. The dynamics is given by the time-dependent decay rate, which reaches its minimal value 0 for an atom sitting at the node and maximal value $2\gamma_0$ for the atom at the antinode of the cavity, a well known result from a quantum cavity electrodynamics. Here, we denoted by $\gamma_0$ the atomic decay rate in free space. We describe the motion of the mirror by a time-dependent function $l(t)$, such that the atom-mirror distance is given by $L-l(t)$ for any time $t$.

The atomic diploe operator in interaction picture reads
\beq
	\hat{\bm d} = \bm d \left(\hat{\sigma}_- e^{-i\omega_a t} + \hat{\sigma}_+ e^{i\omega_a t} \right),
	\label{eq d}
\eeq
\noindent where $\omega_a$ is the atomic transition frequency and $\hat{\sigma}_+=\ket{e}\bra{g},  \hat{\sigma}_-=\ket{g}\bra{e},  \hat{\sigma}_z=\ket{e}\bra{e}-\ket{g}\bra{g}=\hat{\sigma}_{+}\hat{\sigma}_{-}-\hat{\sigma}_{-}\hat{\sigma}_{+}$ are the usual two-level atom operators with a ground and excited states $\ket{g}$ and $\ket{e}$.

The dipole interaction Hamiltonian is equal to the scalar product of the atomic dipole (\eq{\ref{eq d}}) and the electric field (the positive part of which is given by \eq{\ref{eq E+}}), $\hat{H}_I = -\hat{\bm d} \cdot \hat{\bm E}$. The Hamiltonian in the interaction picture, after making the rotating wave approximation, is given by
\beq
	\hat{H}_I(t)=-i \hbar \sum_{\lambda} \int {\rm d} \omega [g_{\omega, \lambda} \hat{\sigma}_{+} \hat{a}_{\omega} \sin\left[k(L-l(t))\right]\, e^{-i(\omega-\omega_a)t}-h.c.].
\eeq
In the following, we assume that the atomic dipole ${\bm d}$ is oriented parallel to the polarization of the field $\bm \epsilon$ and thus yielding the maximized coupling
\beq
g_{\omega} \equiv g_{\omega, \lambda} = \frac{{d}\,A(\omega)}{\hbar},
\eeq
\noindent where is $d=|{\bm d}|$ is the scalar atomic dipole momentum.

The Heisenberg equations of motion of the field and atomic operators are
\beqa
\label{eq_dd ef1}
 \dot{\hat{a}}_{\omega} &=& g^*_{\omega} \sin\left[k(L-l(t))\right]\,e^{i(\omega-\omega_a)t} \,\hat{\sigma}_{-},
\eeqa
 \vskip -0.5cm
\beqa
\label{eq_dd_am1}
 \dot{\hat{\sigma}}_{-} &=& -\frac{\gamma'}{2}\hat{\sigma}_{-} +\hat{\zeta}_-\\ \nonumber
 &+&\hat{\sigma}_{z}\int {\rm d} \omega g_{\omega} \hat{a}_{\omega} \sin\left[k(L-l(t))\right]\,e^{-i(\omega-\omega_a)t},
\eeqa
 \vskip -0.5cm
\beqa
\label{eq_dd_az1}
 \dot{\hat{\sigma}}_{z} &=&  -\gamma'(\hat{\sigma}_{z}+1)+\hat{\zeta}_z \\ \nonumber
 &-& 2 \int {\rm d}  \omega \sin\left[k(L-l(t))\right] \,[ g_{\omega} \hat{\sigma}_{+} \hat{a}_{\omega}\,e^{-i(\omega-\omega_a)t}+h.c.],
\eeqa
\noindent in which the decay term $\gamma'$ and the noise operators $\hat{\zeta}$ are introduced to account for the interaction of the atom with the environment. The explicit form of the noise operator is discussed in \cite{Wang_2011}. Moreover, as a consequence of our initial conditions, the noise operators do not come into play, as explained further in the text.
By integrating Eq.(\ref{eq_dd ef1}), we can separate the field operator into two parts:
\beq
\label{eq_dd_ef2}
\hat{a}_{\omega}(t)=\hat{a}_{\omega}(t_0)+ \int_{t_0}^t {\rm d} t'\, g^*_{\omega}\sin\left[k(L-l(t'))\right]\,e^{i(\omega-\omega_a)t'}\,{\hat{\sigma}_{-}(t')},
\eeq
\noindent where the first term refers to the initial field having evolved freely from $t_0$ to $t$ and the second term is the field created by the atomic dipole during the time period $t-t_0$. These contributions are usually called the ``free field" and the ``source field".

After substituting Eq.(\ref{eq_dd_ef2}) into Eq.(\ref{eq_dd_am1}) and Eq.(\ref{eq_dd_az1}), we get the modified optical Bloch equations
\beqa
\label{eq_dd_am2}
\dot{\hat{\sigma}}_{-}(t) &=& -\frac{\gamma'}{2}\hat{\sigma}_{-}(t) +\hat{\zeta}_- \\ \nonumber
&+& \hat{\sigma}_{z}(t)\int {\rm d} \omega \, g_{\omega}\,\sin\left[k(L-l(t))\right]\,e^{-i(\omega-\omega_a)t}\, \hat{a}_{\omega}(t_0) \\ \nonumber
&+& \hat{\sigma}_{z}(t)\int_{t_0}^{t} {\rm d} t'\,\int {\rm d}  \omega \,{|g_{\omega}|}^2 \hat{\sigma}_{-}(t') \\ \nonumber
&\times&  \sin\left[k(L-l(t))\right] \,\sin[k(L-l(t'))]\,e^{-i(\omega-\omega_a)(t-t')},
\eeqa
\beqa
\label{eq_dd_az2}
\dot{\hat{\sigma}}_{z}(t) &=&  -\gamma'(\hat{\sigma}_{z}(t)+1)+\hat{\zeta}_z \\ \nonumber
&-& 2 \int {\rm d} \omega \sin\left[k(L-l(t))\right] \,\left( g_{\omega} \hat{\sigma}_{+}(t) \hat{a}_{\omega}(t_0) \,e^{-i(\omega-\omega_a)t}+h.c.\right) \\ \nonumber
&-& 2 \int_{t_0}^{t} {\rm d} t'\,\int {\rm d} \omega \,{|g_{\omega}|}^2 \,\sin\left[k(L-l(t))\right] \,\sin[k(L-l(t'))]\\ \nonumber
&\times&\left(e^{-i(\omega-\omega_a)(t-t')} \hat{\sigma}_{+}(t)\hat{\sigma}_{-}(t')+h.c.\right).
\eeqa


\subsection{Characteristics of the quantum memory setup }
We will now focus in the quantum memory application of the considered setup and qualitatively discuss some characteristics the system should meet. One can thus make further assumptions which in turn enables to simplify the above equations.

Let's denote a round-trip time of the light between the atom and the mirror as $\tau = 2L/c$. In the ideal case we wish to absorb a photon by the atom, where the maximum coupling reaches $2\gamma_0$, as discussed earlier in this section, and indicates a relevant timescale (lower limit) for the photon duration. To prevent losses due to spontaneous emission during the write process, we thus require that (i) $\gamma_0 \tau \ll 1$ (Markov approximation). Furthermore, the coupling can be tuned between its maximal and minimal value by changing the atom-mirror distance on the order of the wavelength $\lambda$, thus changing the position of the atom between nodes and antinodes at will. We thus assume that (ii) $l(t) \approx \lambda$. Typically, $c \tau$ can be of the order of many wavelengths, so $\tau \gg l(t)/c$. With these arguments, we neglect the change in the operators on time scales smaller or equal to $\tau$, so that $\hat{\sigma}(t \pm \tau) \approx \hat{\sigma}(t \pm l(t)/c) \approx \hat{\sigma}(t)$. On the other hand, one must keep such dependence in all phases present in the equations in order to preserve the interferences. Then the atomic operators evolve as
\beqa
    \label{eq dd_am3}
	\dot{\hat{\sigma}}_{-}(t) &=& -\gamma(t) \,\hat{\sigma}_{-}(t) +\hat{\zeta}_- \\ \nonumber
	&+& \hat{\sigma}_{z}(t)\int {\rm d} \omega g_{\omega}\,\sin\left[k(L-l(t))\right]\,e^{-i(\omega-\omega_a)t}\, \hat{a}_{\omega}(t_0),
\eeqa
\vskip -0.5cm
\beqa
    \label{eq dd_az3}
	\dot{\hat{\sigma}}_{z}(t) &=& - \gamma^z(t) \, \left(\hat{\sigma}_{z}(t)+1\right)+\hat{\zeta}_z \\ \nonumber
	&-& 2 \int {\rm d} \omega \sin\left[k(L-l(t))\right] \,\left( g_{\omega} \hat{\sigma}_{+}(t) \hat{a}_{\omega}(t_0) \,e^{-i(\omega-\omega_a)t}+h.c.\right).
\eeqa
The time-dependent decay rates $\gamma(t)$ and $\gamma^z(t)$ are functions of the motion of the mirror $l(t)$
\vskip -0.5cm
\beq
    \label{eq_gamma_t}
	\gamma(t) = \frac{\gamma'}{2}+\frac{\gamma_p}{2}\, \left(1-e^{i\omega_a\left(\tau-\frac{2 l(t)}{c}\right)}\right),
\eeq
\vskip -0.5cm
\beqa
    \label{eq_gamma_z_t}
	\gamma^z(t) = \gamma'+ \gamma_p \, \left(1-\cos\left[\omega_a\left(\tau-\frac{2 l(t)}{c}\right)\right] \right)=2 Re [\gamma(t)],
\eeqa
\noindent where $\gamma_p$ is the decay into the pulse mode, which makes up the standard free space decay rate $\gamma_0$ together with the decay into the environment (the non-pulse mode) $\gamma'$, such that $\gamma' + \gamma_p = \gamma_0$. Using the Weisskopf-Wigner theory \cite[p. 207]{Scully_1997}, the explicit formula of $\gamma_p$ is given by $\gamma_p = \pi {|g_{\omega_a}|}^2$. We would like to note that in the derivation of the equations of motion \eqs{\ref{eq dd_am3}-\ref{eq dd_az3}}, various contributions to the level shifts are omitted (Lamb shift, Van der Waals and Casimir-Polder shifts). The reason is that for a typical atom-mirror distance $L \gg \lambda$, these level shifts are either negligible or constant \cite{Hetet_2010}. The only relevant dynamical level shift, which is the imaginary part of $\gamma(t)$ \eq{\ref{eq_gamma_z_t}} is included.

With the general equations for the atomic operators \eqs{\ref{eq dd_am3}-\ref{eq dd_az3}} and the electric field operator, discussed more in detail in Appendix B \eqs{\ref{eq E+ scatt}--\ref{eq E2 scatt}}, it is now possible to study the dynamics of absorption, storage and retrieval of a single-photon wave packet. Since the absorption medium is a two-level system, we will consider in the following the storage process only of a single photon in Fock state \cite[p. 243]{Loudon_2000,Wang_2011}. The single photon Fock state pulse is defined as
\beq
\ket{1_{p}} = \int {\rm d} \omega \,f_p(\omega) \hat{a}^\dagger_{\omega} \ket{0} = \int {\rm d} t \,\xi_{p}(t) \hat{a}^\dagger_{t} \ket{0},
\eeq
\noindent where $f_p(\omega)$ is the spectral distribution function and $\xi_{p}(t)$ is the temporal shape of the wave packet, which are related by Fourier transform
\beq
\xi_{p}(t) =\frac{1}{\sqrt{ 2 \pi}} \int {\rm d} \omega \,f_p(\omega)\,e^{-i(\omega-\omega_0)t}.
\eeq
In the following, we use $p={\it {in},\it {out}}$ in order to label the input and output pulse waveform $\xi_{p}(t)$. Moreover, all the other considered quantities are labeled by {\it w} and {\it r} for the write and read process, respectively.


\subsection{Write process: Absorption}
\label{sec_write}

During the write process, we wish to efficiently absorb the incoming photon and thus maximize the probability $P$ that the atom gets excited, where ideally $P= 1$. Considering an incoming photon which is nonzero only between times $t_w$ and $t_w^0$, which are the start and end time of the write process, the write efficiency is defined as
\beq
    \label{eq eta_w def}
	\eta_w = \frac{P(t_w^0)}{\int_{t_w}^{t_w^0} {\rm d} t\, {|\xi_{in}(t)|}^2}.
\eeq
In the case of a single photon pulse, which satisfies the normalization condition $\int_{t_w}^{t_w^0} {\rm d} t\, {|\xi_{in}(t)|}^2=1$, the write efficiency is then simply $\eta_w = P(t_w^0)$. The excitation probability can be calculated using its definition
\beq
	P(t) = \frac{1}{2}\,\Big( 1 + \bra{\psi(t_w)}\hat{\sigma}_{z}(t)\ket{\psi(t_w)} \Big),
\eeq
\noindent where $\ket{\psi(t_w)}=\ket{g,1_{in},0_e}$ is the initial state of the total system, with the atom being in its ground state, an incident single photon in Fock state and the environment is in the vacuum state.

So far we have included the environmental decay channel described by the decay rate $\gamma'$ and the related noise operators $\hat{\zeta}$. One important point is that when considering the initial state of the environment to be the vacuum state, the noise operators do not come into play, since $\bra{\psi(t_w)} \hat{\zeta} \ket{\psi(t_w)}=0$ (see also the discussion in \cite{Wang_2011}). Although it is very challenging to achieve experimentally, in the following we assume that all modes of the field radiated by the atom to the mirror half-space (i.e. to the left of the atom in \fig{\ref{fig setup}}) are covered by the mirror. This implies $\gamma'=0$, $\gamma_p = \gamma_0$. It also enables us to separate the effect of the time-dependent coupling $\gamma(t)$ from the effect of the decay to the environment. It is then clear from \eq{\ref{eq_gamma_z_t}} that the time-dependent decay rate $\gamma^z(t)$ changes between $[0,2\gamma_0]$ depending on the position of the mirror.

The set of coupled differential equations \eqs{\ref{eq dd_am3},\ref{eq dd_az3}} for the atomic operators gives the absorption probability ({see Appendix A for details})
\beq
    \label{eq Pe write}
	P(t_w^0) = {\left|e^{-\Gamma_w{\left(t_w^0\right)}}\int_{t_w}^{t_w^0} {\rm d} t \,e^{\Gamma_w{(t)}} g_w(t) \,\xi_{in}(t) \right|}^2,
\eeq
where we define
\beq
\Gamma_w{(t)}=\int_{t_w}^{t} {\rm d} t' \gamma_w{(t')},
\eeq
\noindent with $\gamma_w{(t)}$ given by \eq{\ref{eq_gamma_t}} and the subscript {\it w} indicates the write process in order to distinguish it from the read process which has in principle different decay function $\gamma_r{(t)}$ . The effective time-dependent coupling strength reads
\beq
g_w(t)= \sqrt{2 \gamma_0} \, \sin \left[\omega_a\left(\frac{\tau}{2}-\frac{l(t)}{c}\right)\right]=\sqrt{\gamma_w^z(t)}.
\eeq
The goal is now to find the time-dependent $\gamma_w^z(t)$ that maximizes the write efficiency for a given input field $\xi_{in}(t)$, which can be done using Lagrange multiplier optimization \cite[p. 169]{Riley_2006},
\beq
\label{eq lag}
\frac{\delta}{\delta \xi^*_{in}(t)}\left[P(t_w^0)+\lambda \left(\int_{t_w}^{t_w^0} {\rm d} t\, {|\xi_{in}(t)|}^2-1\right)\right]=0
\eeq
\noindent where $\lambda$ is the Lagrange multiplier.
This results in the optimized write efficiency
\beq
\label{eq etaw}
	\eta_w = 1- e^{-\Gamma_w^z(t_w^0)},
\eeq
\noindent with $\Gamma_w^z{(t)}=\int_{t_w}^{t} {\rm d} t' \gamma_w^z{(t')}$, and the time-dependent decay rate satisfying
\beq
\label{eq_gamma_zw}
\gamma_w^z(t)=
\left\{
\begin{array}
    {r@{\quad:\quad}l}
    \frac{\eta_w\,{|\xi_{in}(t)|}^2}{(1-\eta_w)+\eta_w \int_{t_w}^{t_w^0} {\rm d}  t'\, {|\xi_{in}(t')|}^2} & \gamma_w^z(t) \leq 2 \gamma_0;\\
    2 \gamma_0  & \gamma_w^z(t) \geq 2 \gamma_0,
\end{array}
\right.
\eeq
\noindent where we have to account for the physical limitation of the system, $ 0 \leq \gamma_w^z(t) \leq 2\gamma_0$.

After the absorption, the single photon is stored as the excitation of the atom for a time period $T$. During this period, the static mirror position is such that the atom sits at the node, i.e. $\gamma_w^z(t)=0$, so that the atom remains in its excited state, which implies that $P(t_w^0 \leq t \leq t_r^0) = P(t_w^0)$ during the storage period.


\subsection{Read process: Re-emission}
\label{sec_read}
For an on-demand readout of the stored single photon pulse, the atom-light interaction is turned on again at the starting time of the readout process $t_r^0 = t_w + T$. As discussed above, we consider no losses during the storage process, so that $P(t_r^0)=P(t_w^0)=\eta_w$. In analogy to the write efficiency, we define the efficiency of the readout process ending at time $t_r$ as
\beq
    \label{eq eta_r def}
	\eta_r = \frac{\int_{t_r^0}^{t_r} {\rm d}  t\, {|\xi_{out}(t)|}^2}{P(t_r^0)}.
\eeq
The temporal shape of the outgoing pulse $\xi_{out}(t)$ can be derived from the electric field operators \eqs{\ref{eq E+ scatt}--\ref{eq_xiout_def}} in Appendix B as
\beqa
\label{eq_xiout1}
\xi_{out}(z,t) &=&  \sqrt{\frac{2}{\pi}}\, \frac{1}{A(\omega_a)} \bra{\psi_0}\hat{E}_{out}^+(z,t)\ket{\psi(t_r^0)} \\ \nonumber
&=& i \sqrt{\frac{2}{\gamma_0}} \, e^{-i \omega_a (t-z/c+\tau/2)}\,\gamma_r(t) \bra{\psi_0} \hat{\sigma}_{-}(t-z/c) \ket{\psi(t_r^0)},
\eeqa
\noindent with $\ket{\psi_0}=\ket{g,0_{in},0_e}$ and $\ket{\psi(t_r^0)}=\ket{e,0_{in},0_e}$.
The evolution of the atomic operators can be also found using \eqs{\ref{eq dd_am3}-\ref{eq dd_az3}}
\beq
\label{eq_am_r1}
\bra{\psi_0} \hat{\sigma}_{-}(t-z/c) \ket{\psi(t_r^0)} = \sqrt{P(t_r^0)} \,e^{-\Gamma_r(t)}.
\eeq
Since we are interested in the output pulse at certain position $z \geq L$, the temporal shape of the output pulse $\xi_{out}(t)$ reads
\beq
\label{eq_xiout2}
\xi_{out}(t) = \xi_{out}(z,t)\big |_{z=D \geq L} = i \sqrt{\frac{2\,P(t_r^0)}{\gamma_0}} \, e^{-i \omega_a (t-D/c+\tau/2)}\,\gamma_r(t)\,e^{-\Gamma_r(t)},
\eeq
\noindent which implies that the temporal shape of the output pulse can be adjusted by controlling the time-dependent read decay rate
\beq
	\gamma^z_r(t)=\frac{{|\xi_{out}(t)|}^2}{\eta_w - \int_{t_r^0}^{t} d t'\, {|\xi_{out}(t')|}^2},
\eeq
which is again subjected to the constraint that $ 0 \leq \gamma_r^z(t) \leq 2\gamma_0$.
Plugging \eq{\ref{eq_xiout2}} into \eq{\ref{eq eta_r def}} one finds the expression for the read efficiency
\beq
\label{eq etar}
\eta_r = 1- e^{-\Gamma_r^z(t_r^0)},
\eeq
\noindent with $\Gamma_r^z{(t)}=\int_{t_r^0}^{t} {\rm d} t' \gamma_r^z{(t')}$.

The total quantum memory efficiency is given by
\beq
\eta=\eta_w\,\eta_r=(1- e^{-\Gamma_w^z(t_w^0)})\,(1- e^{-\Gamma_r^z(t_r^0)}).
\eeq
So far we have derived an expression for the efficiency of the readout process as a function of a time-dependent readout decay rate $\gamma^z_r (t)$. We should however emphasize a simple reflection that, an atom in the excited state with a nonzero coupling to the field will necessarily decay. Typically, for a constant $\gamma_r$, the decay will be exponential with $\eta_r$ approaching 1 already for times of $1/\gamma_r$. The readout can be thus made simply by ``waiting".

In the following, we would rather require that the quantum memory device yields the maximum fidelity $F =1$.
The memory fidelity is expressed in terms of the outgoing pulse's projection on the input pulse as
\beqa
\label{eq F def}
F = \left| \bra{1_{in}} 1_{out} \rangle \right|^2 = \frac{{\left|\int d t \, \xi^*_{in}(t)\xi_{out}(t)\right|}^2}{\int d t \, {\left|\xi_{in}(t)\right|}^2 \cdot \int d t \,{\left|\xi_{out}(t)\right|}^2}.
\eeqa
Obviously, the ideal fidelity is achieved when the output pulse has the same shape as the input pulse $\xi_{out}(t)= \sqrt{\eta} \,\xi_{in}(t-T)$, which can be achieved by changing the read decay rate in the following way,
\beq
\label{eq_gamma_zr}
\gamma_r^z(t)=
\left\{
\begin{array}
    {r@{\quad:\quad}l}
    \frac{\eta_r\, {|\xi_{in}(t+(t_r^0-t_w^0))|}^2}{1 - \eta_r \,\int_{t_r^0}^{t} d t'\, {|\xi_{in}(t'+(t_r^0-t_w^0))|}^2} & \gamma_r^z(t) \leq 2 \gamma_0;\\
    2 \gamma_0  & \gamma_r^z(t) \geq 2 \gamma_0.
\end{array}
\right.
\eeq
Due to the similarity of the underlying physics, we would like to note that the expressions for read and write efficiency \eqs{\ref{eq etaw},\ref{eq etar}} are analogous to those in Ref. \cite{Green_2011}.


\section{Simulations and possible implementations}
\label{sec Sim_Imp}
\subsection{Simulation with time-bin qubit}
\label{sec Simulation}
\begin{figure}[h!]
\includegraphics[scale=0.3]{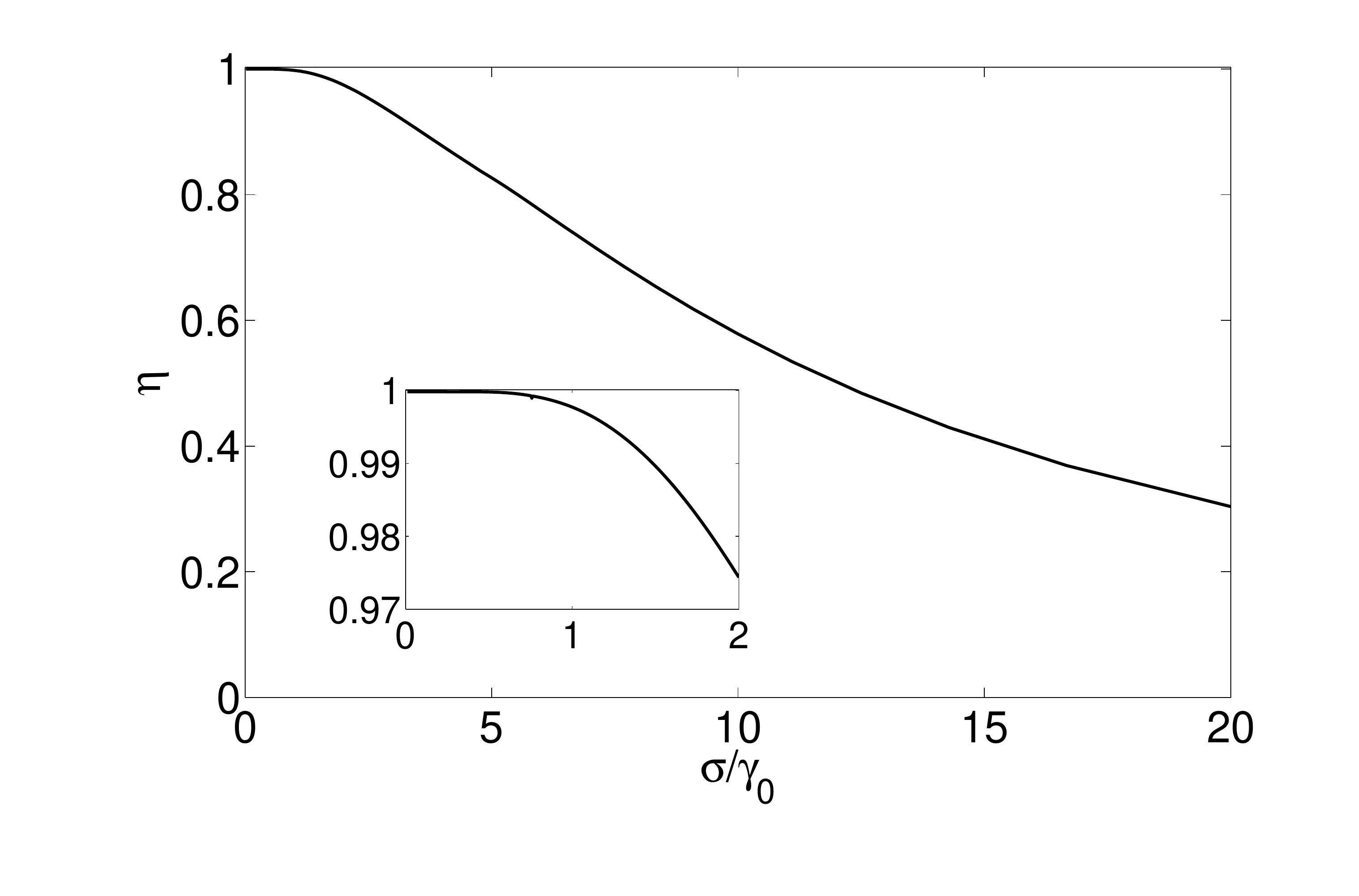}
\caption{ Total efficiency of the quantum memory device as a function of the bandwidth $\sigma$ of the input pulse. The curve was obtained with Gaussian-shaped time-bin single photon wave packet \eq{\ref{eq_tb}} for ideal fidelity $F=1$. The inset is a zoom of the region of $\sigma/\gamma_0$ between $0$ and $2$.}
\label{fig eta}
\end{figure}

With the help of \eqs{\ref{eq_gamma_zw},\ref{eq_gamma_zr}}, we can now study the performance of the quantum memory as a function of the input light field. In the following, we consider a specific case of a normalized Gaussian-shaped time-bin single photon pulse described as
\beq
\label{eq_tb}
\xi_{in}(t) =  \alpha\,e^{-\frac{{(t-t_{1})}^2 \sigma^2}{2}} + \beta\, e^{i \phi}\,e^{-\frac{{(t-t_{2})}^2 \sigma^2}{2}},
\eeq
\begin{figure}[h!]
\begin{minipage}{1 \linewidth}
\includegraphics[scale=0.32]{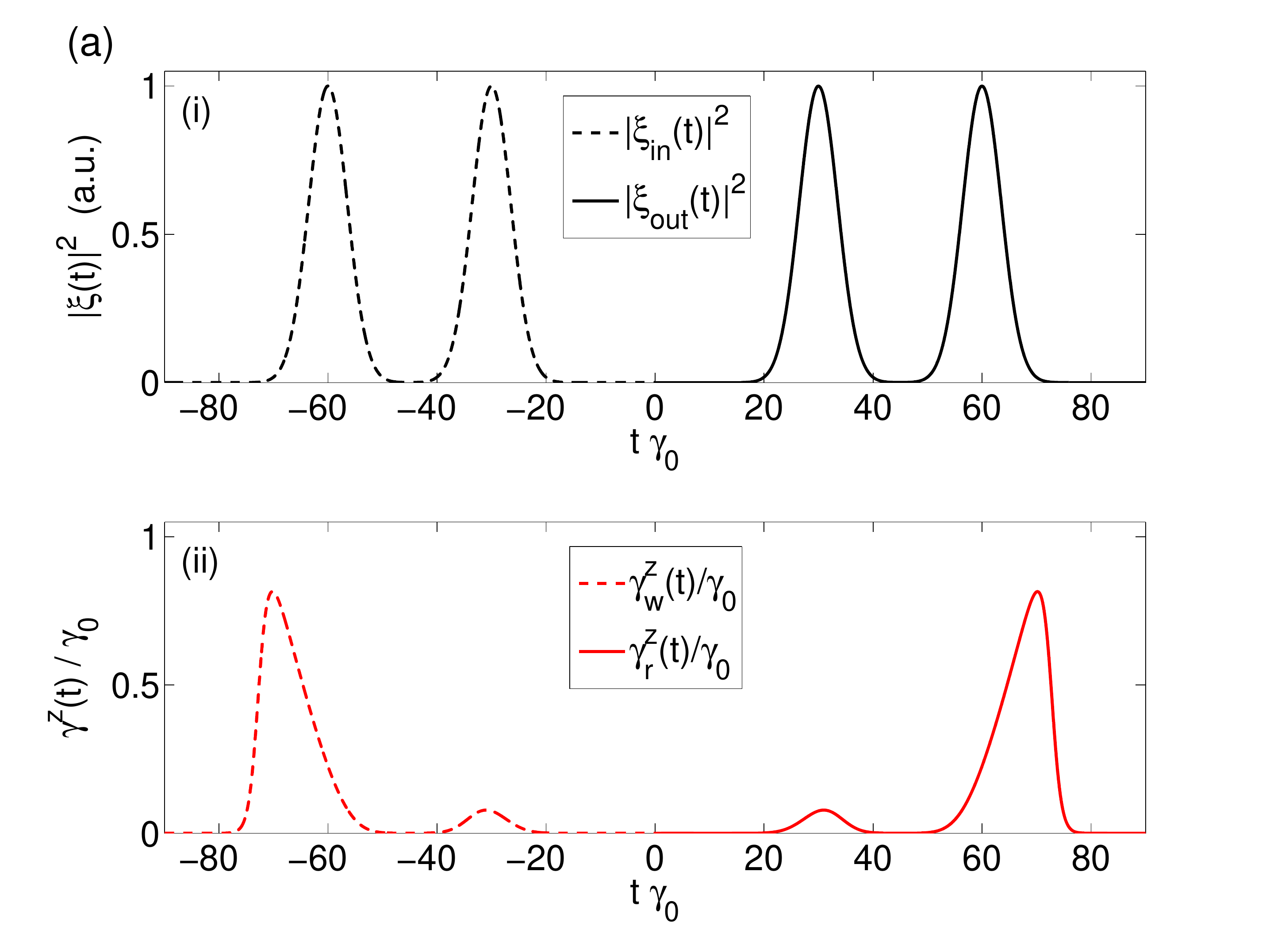}
\vspace{0.0cm}
\end{minipage}
\begin{minipage}{1 \linewidth}
\includegraphics[scale=0.32]{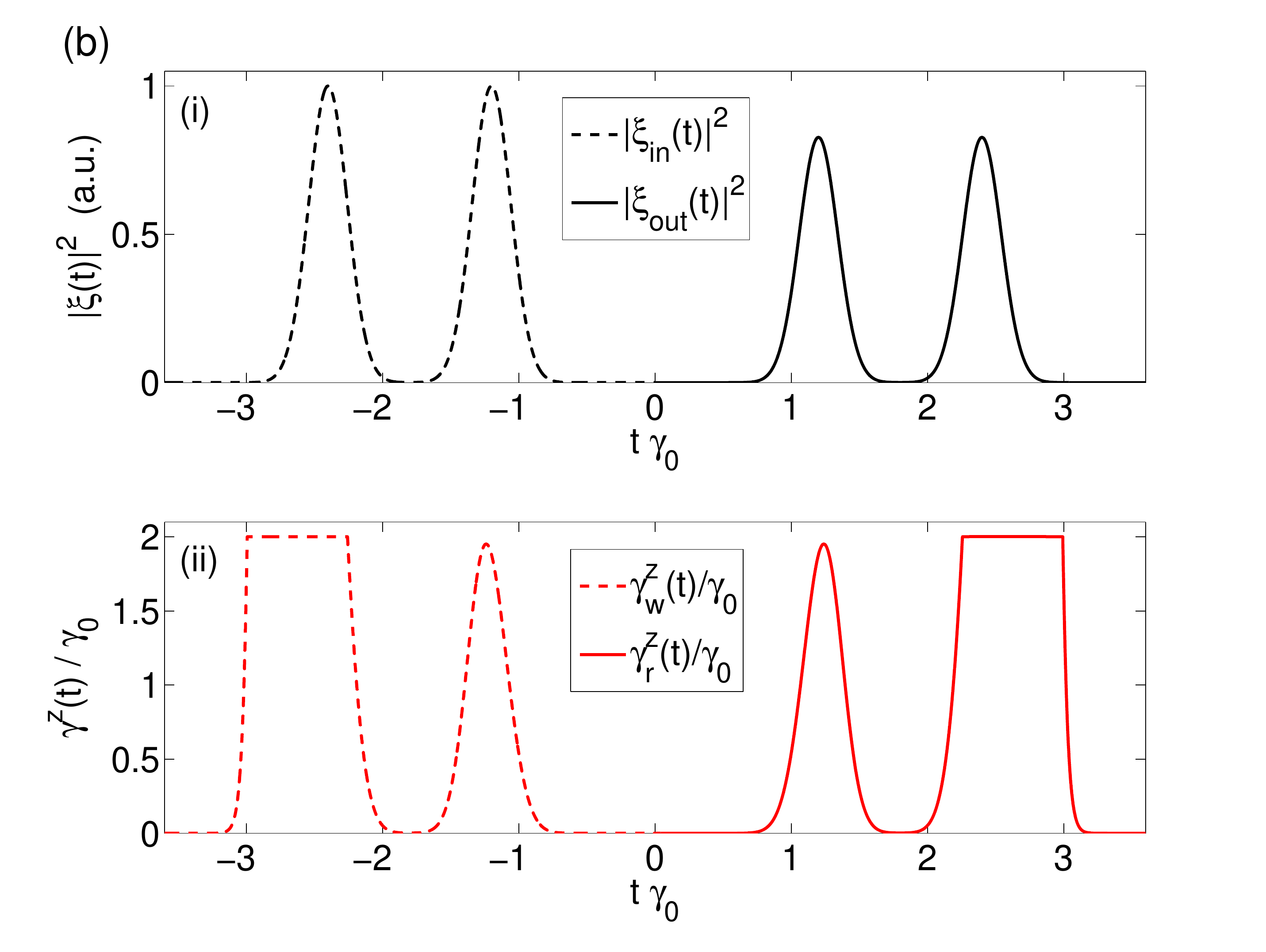}
\vspace{0.0cm}
\end{minipage}
\caption{(Color online) Storage ($t<0$) and retrieval ($t>0$) of a Gaussian-shaped time-bin single photon pulse for different values of bandwidth: (a) $\sigma =0.2 \gamma_0$ ;  (b) $\sigma=5 \gamma_0$. The input intensity (dashed black line) and the output intensity (solid black line) of the pulse is shown in (i), with the input intensity normalized to amplitude 1. The required optimum write and read decay rate $\gamma^z_w(t)$ and $\gamma^z_r(t)$ is shown by dashed and solid red line in (ii), respectively. It can be seen that for the smaller bandwidth, case (a), $ {\it Max} [\gamma^z_{w,r}(t)] < 2 \gamma_0$ and the efficiency is close to $1$; on the other hand, for the larger bandwidth, case (b), where $\gamma$ has to be truncated at $2 \gamma_0$, the efficiency is less than $1$. }
\label{fig inout}
\end{figure}

\noindent where the real coefficients $\alpha,\,\beta$ satisfy $\alpha^2 + \beta^2 =1$, $t_{2}-t_{1}$ is the relative time delay, $\phi$ is the relative phase between the two time bins and the bandwidth $\sigma$ is assumed the same for each time bin. The performance of the quantum memory is studied for different bandwidths $\sigma$ of the pulse with $\alpha=\beta$. In \fig{\ref{fig inout}}, two particular situations are considered, one with photon bandwidth smaller and the other one with photon bandwidth larger than the double of the atomic decay rate $2 \gamma_0$. In \fig{\ref{fig inout}} (a), we set $\sigma =0.2 \gamma_0$. In this case, the quantum memory efficiency reaches its maximal value, $\eta =1$: the amplitude of the output pulse (solid black line) is the same as the input pulse (dashed black line) as can be seen from \fig{\ref{fig inout}} (a)(i). On the other hand, \fig{\ref{fig inout}} (b) with $\sigma =5 \gamma_0$ shows a decrease of the efficiency. The optimized decay rates $\gamma^z_w(t)$ and $\gamma^z_r(t)$ are represented by dashed and solid red lines respectively. The shapes of the optimum coupling decay rates are given by \eqs{\ref{eq_gamma_zw},\ref{eq_gamma_zr}} and might be qualitatively understood as follows. For write efficiencies $\eta_w \approx 1$, the write decay rate $\gamma^z_w(t)$ is proportional to the intensity divided by the time integral of the intensity. This ratio can be high at the beginning of the write process (first time bin), when the denominator is small, but gets significantly smaller for the second time bin. Similar argument holds for the read coupling decay. It is possible to plot the motion of the mirror $l(t)$ instead of the coupling decay rate $\gamma^z(t)$ (see \eq{\ref{eq_gamma_z_t}}). In the example presented in \fig{\ref{fig inout}}, the motion of the mirror is similar to the coupling decay rate with the $l(t)$ ranging from 0 to $\lambda/4$ (corresponding to $2\gamma_0$ for the decay rate) and we do not plot it explicitly. Finally, one can see that for the photon bandwidth larger than the cutoff frequency of the system $2\gamma_0$, the optimum decay rates $\gamma^z_{w,r}(t)$ exceed this cutoff and are thus truncated at $2\gamma_0$. This results in the decrease of the storage efficiency, as shown in \fig{\ref{fig inout}} (b)(i). The storage efficiency as a function of the ratio between the photon bandwidth and the atomic decay rate is shown in \fig{\ref{fig eta}}. The efficiency starts to decrease for $\sigma/\gamma_0 \approx 0.85$ which corresponds to $ {\it FWHM } =2 \sqrt{2{\rm Log}2} \,\sigma = 2 \gamma_0$, as expected.

\subsection{Implementations}
\label{sec Implementations}

We will now discuss possible implementations of our protocol. The described quantum memory device requires a single two-level system with a tunable distance to the mirror and a strong coupling to the light field. Strongly coupled two-level systems can be implemented using optical setups with ions and atoms \cite{Eschner_2001, Tey_2008}, quantum dots \cite{zhang_2008}, superconducing qubits in circuit QED configuration \cite{Devoret_2004, You_2011} or atoms coupled to surface plasmons on conducting nanowires \cite{Chang_2007} or to tapered optical nanofibers \cite{Nayak_2011}.

As for the quantum optical implementations, there is a variety of atoms and ions used in trapping experiments, typical examples being e.g. a $^{138}{\rm Ba}^+$ ion in a Paul trap \cite{Eschner_2001, Slodicka_2010, Hetet_2011} or $^{87}{\rm Rb}$ atom in a FORT trap \cite{Tey_2008}. In the case of ${\rm Ba}^+$ ions, the typical two-level transition is between the ground state $\ket{6S_{1/2},m_F=1/2}$ and excited state $\ket{6P_{1/2},m_F=-1/2}$ at $\lambda=493$ nm with a linewidth $\gamma_0 = 15$ MHz \cite{Slodicka_2010}. For this system, as experimental realization with half-cavity an tunable atom-mirror has been reported \cite{Dubin_2007} (an analogous experimental setup with quantum dot has been also realized \cite{zhang_2008} ). This, together with an atom-mirror distance $L$ of order of centimeters, meets very well the assumption required for quantum memory: $\gamma_0\tau \ll 1$. On the other hand, the durations of incoming photon of the order up to $1/\gamma_0$ require the motion of the mirror at the same time scale, which might be hard to achieve by a mechanical motion. One possible solution is to use a long-lived quadrupole transition (for which the lifetime can be seconds (e.g. Ca$^{+}$ or Ba$^{+}$)) which would allow for slower mechanical motion of the mirror achievable with current technology. Another possibility is to move the atom itself, which can be done very fast in Dipole or Paul traps. The drawback of this approach is that the atom would get slightly out of the focus of the mirror, reducing thus the maximum achievable coupling decay rate \cite{Hetet_2010}.
It might be also be possible to use an EOM in the integrated setup to modulate the optical path-length \cite{Chang_2007, Nayak_2011}.

The spatial overlap of the incident field and the atomic dipole pattern needs to be taken into account in realistic systems, as discussed elsewhere in more detail for hemispherical mirror \cite{Hetet_2010} and for parabolic mirror \cite{Sondermann_2008, Stobinska_2009}. The consequence of imperfect spatial overlap is the decay into the environment $\gamma'$ which would reduce the write efficiency as well as, and more importantly, the storage process (since the storage time $T$ is often required to be much larger than the photon duration, the population of the excited state $\propto {\rm exp}(-\gamma' T)$ is more affected during the storage, because ${\rm exp}(-\gamma' T) \ll {\rm exp}(-\gamma't_p)$, where $t_p$ is the pulse duration). Obviously, the quantum memory scheme works only for single photon Fock states, which are available experimentally \cite{Kuhn_2002}. Finally, we would like to mention that the quantum memory works also for the polarization qubits. In this case the required level scheme is a V configuration, standardly available for typical atoms used in the experiments.

The proposed quantum memory device can be also implemented in the fast growing domain of circuit QED, where the effective two-level system can be realized by different kinds of superconducting qubits \cite{Devoret_2004, You_2011}. Typical resonant frequencies of a superconducting qubit lay in the microwave region of order of 1-10 GHz with population decay rates of order of 1-10 MHz \cite{Wallraff_2004, Houck_2007, Abdumalikov_2010}. Generation of various photonic states, including a single photon Fock state, was demonstrated in several experiments \cite{Houck_2007, Hofheinz_2008, Hofheinz_2009} laying thus the ground for potential realization of the presented quantum memory scheme. The configuration of superconducting qubits coupled to a transmission line resonator has the beauty of well defined one-dimensional (1D) mode and perfect spatial overlap, which results in strong atom-light interaction. Moreover, an open transmission line with one side terminated by a SQUID operated with a variable magnetic flux, acts as a mirror with a tunable qubit-mirror distance. This was realized recently in the remarkable demonstration of dynamical Casimir effect by Wilson {\it et al.} \cite{Wilson_2011}, with oscillation frequency of the SQUID mirror of 11 GHz. Currently, schemes and proposals directly linked to the quantum memory applications are actively investigated both theoretically \cite{Marcos_2010} and experimentally \cite{Schuster_2010, Wu_2010, Mariantoni_2011_nphy, Mariantoni_2011_sci}. In one of the realized experiments, a superconducting qubit with a large decoherence rate (order of MHz) was coupled to a transition with a long coherence time (up to 2 ms) in a NV center in a diamond \cite{Kubo_2010}. This technique can be applied also to our proposal to achieve long storage time for the microwave photons.


\section{Conclusion}
\label{sec Conclusion}

In conclusion, we showed by a fully quantized calculation, that a single photon Fock state pulse with various temporal shapes can be efficiently stored and retrieved from a quantum memory device consisting of a single two-level atom in a half cavity. The principle is that the time-dependent atomic decay rate can be dynamically tuned between zero and the maximum $2 \gamma_0$ by changing the distance between the atom and the mirror. The cutoff frequency of the system, given by double of the free space decay rate of the atom, imposes the limits on the input photon bandwith for which the photon can be efficiently stored. We analyzed the dependence of the storage efficiency as a function of the photon bandwidth. Finally, we discussed possible implementations of the proposed quantum memory scheme, such as single atoms/ions in a half cavity or a superconducting qubit coupled to a 1D transmission line terminated by a SQUID.
~\\

\section{Acknowledgements}

We would like to thank Colin Teo, Jing Yan Haw and Luk\'a\v{s} Slodi\v{c}ka for useful discussions. This work was supported by the National Research Foundation and the Ministry of Education, Singapore. G. H. acknowledges support by a Marie Curie Intra-European Action of the European Union.


\bibliographystyle{prsty}
\bibliography{qo_ref}

\begin{thebibliography}{10}

\bibitem{Duan_2001}
L.~M. Duan, M.~D. Lukin, J.~I. Cirac, and P. Zoller, Nature {\bf 414},  413
  (2001).

\bibitem{Sangouard_2009b}
N. Sangouard, C. Simon, H. de~Riedmatten, and N. Gisin, Rev. Mod. Phys. {\bf
  83},  33  (2011).

\bibitem{Hammerer_2010}
K. Hammerer, A.~S. S�rensen, and E.~S. Polzik, Rev. Mod. Phys. {\bf 82},
  1041  (2010).

\bibitem{Simon_2010}
C. Simon {\it et~al.}, Eur. Phys. J. D {\bf 58},  1  (2010).

\bibitem{Specht_2011}
H.~P. Specht {\it et~al.}, Nature {\bf 473},  190  (2011).

\bibitem{Stute_2011}
A. Stute {\it et~al.}, arXiv:1105.0579v1  (2011).

\bibitem{Yilmaz_2010}
R. Yano, M. Mitsunaga, and N. Uesugi, Phys. Rev. Lett {\bf 105},  033601
  (2010).

\bibitem{Devoret_2004}
M.~H. Devoret, A. Wallraff, and J.~M. Martinis, arXiv:0411174v1  (2004).

\bibitem{Dutt_2007}
M.~V.~G. Dutt {\it et~al.}, Science {\bf 1},  1312  (2007).

\bibitem{Tey_2008}
M.~K. Tey {\it et~al.}, Nat. Phys. {\bf 4},  924  (2008).

\bibitem{Tey_2009}
M.~K. Tey {\it et~al.}, New J. Phys. {\bf 11},  043011  (2009).

\bibitem{Chen_2011}
X.~W. Chen, S. Goetzinger, and V. Sandoghdar, arXiv:1106.3024v2  (2011).

\bibitem{Wilk_2010}
T. Wilk {\it et~al.}, Phys. Rev. Lett. {\bf 104},  010502  (2010).

\bibitem{Isenhower_2010}
L. Isenhower {\it et~al.}, Phys. Rev. Lett. {\bf 104},  010503  (2010).

\bibitem{Benhelm_2008}
J. Benhelm, G. Kirchmair, C.~F. Roos, and R. Blatt, Nature Physics {\bf 4},
  463   (2008).

\bibitem{Eschner_2001}
J. Eschner, C. Raab, F. Schmidt-Kaler, and R. Blatt, Nature {\bf 413},  495
  (2001).

\bibitem{Dorner_2002}
U. Dorner and P. Zoller, Phys. Rev. A {\bf 66},  023816  (2002).

\bibitem{Wilson_2003}
M.~A. Wilson {\it et~al.}, Phys. Rev. Lett. {\bf 91},  213602  (2003).

\bibitem{Glaetzle_2010}
A.~W. Glaetzle {\it et~al.}, Optics Communications {\bf 283},  758  (2010).

\bibitem{Green_2011}
A. Green {\it et~al.}, arXiv:1106.3513v1  (2011).

\bibitem{Blow_1990}
K.~J. Blow, R. Loudon, S.~J.~D. Phoenix, and T.~J. Shepherd, Phys. Rev. A {\bf
  42},  4102  (1990).

\bibitem{Domokos_2002}
P. Domokos, P. Horak, and H. Ritsch, Phys. Rev. A {\bf 65},  033832  (2002).

\bibitem{Wang_2011}
Y. Wang, J. Min\'{a}\v{r}, L. Sheridan, and V. Scarani, Phys. Rev. A {\bf 83},
  063842  (2011).

\bibitem{Scully_1997}
M.~O. Scully and M.~S. Zubairy, {\em Quantum Optics} (Cambridge University
  Press, Cambridge, 1997).

\bibitem{Hetet_2010}
G. H\'etet {\it et~al.}, Phys. Rev. A {\bf 82},  063812  (2010).

\bibitem{Loudon_2000}
R. Loudon, {\em The Quantum Theory of Light} (Oxford University Press, Oxford,
  2000).

\bibitem{Riley_2006}
K.~F. Riley, M.~P. Hobson, and S.~J. Bence, {\em Mathematical Methods for
  Physics and Engineering} (Cambridge University Press, Cambridge, 2006).

\bibitem{zhang_2008}
Y. Zhang, V.~K. Komarala, C. Rodriguez, and M. Xiao, Phys. Rev. B {\bf 78},
  241301  (2008).

\bibitem{You_2011}
J.~Q. You and F. Nori, Nature {\bf 474},  589  (2011).

\bibitem{Chang_2007}
D.~E. Chang, A.~S. S{\o}rensen, E.~A. Demler, and M.~D. Lukin, Nature Physics
  {\bf 3},  807   (2007).

\bibitem{Nayak_2011}
K.~P. Nayak {\it et~al.},  in {\em Quantum Electronics and Laser Science
  Conference} (Optical Society of America, ADDRESS, 2011), p.\ QFC2.

\bibitem{Slodicka_2010}
L. Slodi\ifmmode~\check{c}\else \v{c}\fi{}ka {\it et~al.}, Phys. Rev. Lett.
  {\bf 105},  153604  (2010).

\bibitem{Hetet_2011}
G. H\'etet, L. Slodi\ifmmode~\check{c}\else \v{c}\fi{}ka, M. Hennrich, and R.
  Blatt, Phys. Rev. Lett. {\bf 107},  133002  (2011).

\bibitem{Dubin_2007}
F. Dubin {\it et~al.}, Phys. Rev. Lett. {\bf 98},  183003  (2007).

\bibitem{Sondermann_2008}
M. Sondermann, N. Lindlein, and G. Leuchs, arXiv:0811.2098v3  (2008).

\bibitem{Stobinska_2009}
M. Stobi\'{n}ska and R. Alicki, arXiv:0905.4014v1  (2009).

\bibitem{Kuhn_2002}
A. Kuhn, M. Hennrich, and G. Rempe, Phys. Rev. Lett. {\bf 89},  067901  (2002).

\bibitem{Wallraff_2004}
A. Wallraff {\it et~al.}, Nature {\bf 431},  162  (2004).

\bibitem{Houck_2007}
A.~A. Houck {\it et~al.}, Nature {\bf 449},  328  (2007).

\bibitem{Abdumalikov_2010}
A.~A. Abdumalikov {\it et~al.}, Phys. Rev. Lett. {\bf 104},  193601  (2010).

\bibitem{Hofheinz_2008}
M. Hofheinz {\it et~al.}, Nature {\bf 454},  310  (2008).

\bibitem{Hofheinz_2009}
M. Hofheinz {\it et~al.}, Nature {\bf 459},  546  (2009).

\bibitem{Wilson_2011}
C.~M. Wilson {\it et~al.}, Nature {\bf 479},  376  (2011).

\bibitem{Marcos_2010}
D. Marcos {\it et~al.}, Phys. Rev. Lett. {\bf 105},  210501  (2010).

\bibitem{Schuster_2010}
D.~I. Schuster {\it et~al.}, Phys. Rev. Lett. {\bf 105},  140501  (2010).

\bibitem{Wu_2010}
H. Wu {\it et~al.}, Phys. Rev. Lett. {\bf 105},  140503  (2010).

\bibitem{Mariantoni_2011_nphy}
M. Mariantoni {\it et~al.}, Nature Physics {\bf 7},  287–293  (2011).

\bibitem{Mariantoni_2011_sci}
M. Mariantoni {\it et~al.}, Science {\bf 334},  61  (2011).

\bibitem{Kubo_2010}
Y. Kubo {\it et~al.}, Phys. Rev. Lett. {\bf 105},  140502  (2010).

\end{thebibliography}


\renewcommand{\theequation}{A-\arabic{equation}}
\setcounter{equation}{0}  

\section*{Appendix}
\label{sec Appendix}

\subsection{Write process: Absorption}
In order to find out the value of $\bra{\psi(t_w)}\hat{\sigma}_{z}(t)\ket{\psi(t_w)}$ and thus the absorption probability $P(t)$, we have to solve a set of time-dependent differential equations, which is obtained from the average of \eq{\ref{eq dd_az3}} on the initial state $\ket{\psi(t_w)}=\ket{g,1_{in},0_e}$,

\beq
    \label{eq set_w}
	\dot{\bf s}(t) = M\, {\bf s}(t)+{\bf b}.
\eeq
with
\[
	{\bf s}(t)= \left(
	\begin{array}{ccc}
			\bra{g,1_{in},0_e}\hat{\sigma}_{z}(t)\,\ket{g,1_{in},0_e} \\
			\bra{g,1_{in},0_e}\,\hat{\sigma}_{+}(t)\,\ket{g,0_{in},0_e} \\
			 \bra{g,0_{in},0_e}\,\hat{\sigma}_{-}(t)\,\ket{g,1_{in},0_e}
	\end{array} \right)
\]

\beqa
	M = \left(
	\begin{array}{ccc}
			-\gamma^z_w(t) & -2g_w(t) & -2g_w(t) \\
			0 & -\gamma^*_w(t) & 0 \\
			0 & 0 & -\gamma_w(t)
	\end{array} \right),
\qquad
	{\bf b} = \left(
	\begin{array}{ccc}
			-\gamma^z_w(t) \\
			-g_w(t) \\
			-g_w(t)
	\end{array} \right),
\eeqa
with initial condition
\[
	{\bf s}^T(t_0)= \left(
	\begin{array}{ccc}
			-1 & 0 & 0
	\end{array} \right).
\]
The solution of the \eq{\ref{eq set_w}} gives us the absorption probability as a function of time
\beq
	\label{eq Pe write}
    P(t) = {\left|e^{-\Gamma_w{\left(t\right)}}\int_{t_w}^{t} {\rm d} t' \,e^{\Gamma_w{(t')}} g_w(t') \,\xi_{in}(t') \right|}^2.
\eeq

We want to optimize the absorption probability $P$ with respect to the normalized input pulse $\int {\rm d} t\, {|\xi_{in}(t)|}^2=1$. Following the method of Lagrange multipliers and performing a functional differentiation of \eq{\ref{eq lag}}, we find out that the optimum write process must satisfy
\beq
\label{eq_xiin}
\left|\xi_{in}(t)\right| = \frac{1}{\sqrt{\eta_w}} \,e^{-\frac{\Gamma^w_z(t_w^0)}{2}} g^w(t)\,e^{-\frac{\Gamma^w_z(t_w)}{2}}.
\eeq
From here we can extract the required time-dependent decay rate $\gamma_w^z(t)$, which is expressed in \eq{\ref{eq_gamma_zw}}.
~\\


\subsection{Read process: Re-emission}

\renewcommand{\theequation}{B-\arabic{equation}}
\setcounter{equation}{0}  

By inserting the \eq{\ref{eq_dd_ef2}} into the field operator \eq{\ref{eq E+}}, and again under the assumptions $ \gamma_0 \tau \ll 1$ and $l(t) \approx \lambda$, we have the simplified scattered field operator
\beq
\label{eq E+ scatt}
	\hat{E}_{out}^{(+)}(z,t) = \hat{E}_1^{(+)}(z,t) + \hat{E}_2^{(+)}(z,t)
\eeq
where $\hat{E}_1^{(+)}(z,t) $ is the free evolution electric field
\beq
\label{eq E1 free}
	\hat{E}^{(+)}_1(z,t) = i \int_0^{\infty} {\rm d} \,\omega A(\omega)\, \sin[k(z-l(t))]\,e^{-i\omega t}\,a_{\omega}(t_0),
\eeq
and $\hat{E}_2^{(+)}(z,t) $ is the electric field scattered by the atom
\beqa
\label{eq E2 scatt}
	\hat{E}^{(+)}_2(z,t) &=& -  i \frac{\pi}{2} A(\omega_a)\, g_{\omega_a} \\ \nonumber
	&\times&\,\Bigg( e^{-i \omega_a \left(t-\left(\frac{\tau}{2}+\frac{z-2 l(t)}{c}\right)\right)} \,\hat{\sigma}_{-}(t-\frac{z}{c}) \, \Theta (t-\frac{z}{c})
	\\ \nonumber
	&-& e^{-i \omega_a \left(t-\left(\frac{z}{c}-\frac{\tau}{2}\right) \right )}\, \hat{\sigma}_{-}(t-\frac{z}{c}) \, \Theta (t-\frac{z}{c})\,\Theta (\frac{z-L}{c})
	\\ \nonumber
	&-& e^{-i \omega_a \left(t-\left(\frac{\tau}{2}-\frac{z}{c}\right) \right )}\, \hat{\sigma}_{-}(t+\frac{z}{c}) \, \Theta (t+\frac{z}{c})\,\Theta (\frac{L-z}{c})\Bigg ).
\eeqa
In \eq{\ref{eq E2 scatt}}, the usage of the Weisskopf-Wigner theory allows us to put $A(\omega)\approx A(\omega_a)$ out of the integration.
Since we are only interested in the right propagating field in the region $z>L$ (see \fig{\ref{fig setup}}), the step function $\Theta (\frac{L-z}{c})$ implies that the third term in \eq{\ref{eq E2 scatt}} does not contribute to the total field.

To find out the temporal shape of the output pulse after the readout process, we first study the c-number electric field of the input pulse
\beq
\bra{0}\hat{E}^{(+)}_1(z,t)\ket{1_{in}} = -\sqrt{\frac{\pi}{2}} A(\omega_0) \left[\xi(t-z/c)-\xi(t+z/c)\right],
\eeq
\noindent where $\omega_0$ is the carrier frequency of the input pulse, and again $A(\omega)\approx A(\omega_0)$ in Weisskopf-Wigner approximation. When the atom and pulse are in resonance with each other $A(\omega_0)=A(\omega_a)$, the total field is given by the interference of the right propagating pulse $\xi(t-z/c)$ and the left propagating pulse $\xi(t+z/c)$.

Similarly, the electric field contributing to the output pulse at the position of interest (i.e. outside the atom-mirror system, $z>L$), is only the right propagating part. In this case, we have the temporal shape of the output pulse given by
\beqa
\label{eq_xiout_def}
\xi_{out}(z,t) &=&  \sqrt{\frac{2}{\pi}}\, \frac{1}{A(\omega_a)} \bra{\psi_0}\hat{E}_{out}^+(z,t)\ket{\psi(t_r^0)}.
\eeqa
\end{document}